\documentstyle[prl,aps,psfig]{revtex}

\bibstyle{unsrt}

\tighten
\begin{document}
\draft

\preprint{}
\twocolumn[\hsize\textwidth\columnwidth\hsize\csname 
@twocolumnfalse\endcsname

\title{Vortex String Formation in a 3D $U(1)$ Temperature Quench}
\author{Nuno D. Antunes$^1$, Lu\'{\i}s M. A.  Bettencourt$^2$ and 
Wojciech H. Zurek$^2$}
\address{$^1$Theoretical Physics Group,
University of Geneva, Geneva, Switzerland}
\address{$^2$Theoretical Division MS B288, 
Los Alamos National Laboratory, Los Alamos NM 87545}

\date{\today}
\maketitle

\begin{abstract}
We report the first large scale numerical study of 
the dynamics of the second order 
phase transition of a $U(1)$ $\lambda 
\phi^4$ theory in three spatial dimensions. 
The transition is induced by a time-dependent 
temperature drop in the heat bath to which the fields are coupled. 
We present a detailed account of the dynamics of the fields and vortex
string formation as a function of the quench rate. 
The results are found in good agreement to the theory of defect formation
proposed by Kibble and Zurek.
\end{abstract}


\vskip2pc]

Topological defects are of fundamental importance to 
the (thermo)dynamics of phase transitions in some of the 
most fascinating materials in the laboratory 
- eg. superfluids, type II superconductors, liquid crystals - 
and presumably also to symmetry breaking phase transitions in the early 
universe \cite{VS}. When present at low energies they constitute the 
last traces of disorder inherited from high temperatures and 
thus determine the non-equilibrium evolution of many systems.

The theory of defect formation combines the realization, due to 
Kibble \cite{K76} of the paramount role of causality, with the calculation,
due to one of us \cite{Z85}, of the actual size of the 
causally independent domains in a second order phase transition.
Experiments in liquid crystals \cite{LiqC} and in $^3He$ \cite{He3} 
lend support to the resulting theory of the dynamics of second order phase 
transitions.
The evidence from $^4He$ experiments \cite{He4} is more ambiguous while
superconductor and Bose-Einstein condensates may offer exciting future 
possibilities \cite{AngZur}.

Laboratory experiments were, so far unable to test the key 
theoretical prediction - the scaling of the initial density of defects with 
the rate at which the phase transition takes place. Numerical studies 
carried out to date were limited to 1 and 2D systems, where they 
have confirmed the scalings predicted by the theory \cite{LYZ}. 
Moreover, they have focused on transitions induced by 
the explicit change of the mass term in the potential 
which governs dynamics of 
the order parameter field $\langle \phi \rangle$, while the temperature 
$T$ of the heat bath 
to which $\phi$ is coupled was kept constant and relatively 
small. This simplification 
is not necessarily unrealistic - phase transitions in 
$^4He$ can be induced without significant changes in temperature, purely 
by changing pressure. However such 'pressure quenches' are far less common 
than temperature quenches, and temperatures are rarely small. 
Moreover, the analytic approach to 
the pressure quench problem seems to be within reach, although only in 
the limit of very small temperatures. Even in this limiting case 
unanimous conclusion has not been reached \cite{KR,Dz,Lythe}, 
further obviating the need to study quenches numerically.  

Indeed the emerging understanding of the defect production could be 
called into question on the grounds that the present estimates completely 
ignore the inevitable Ginzburg regime. There, just below $T_c$, 
fluctuations can rearrange large spatial regions, which could 
destroy proto-defects produced by  freeze-out \cite{Z85}, or create them at 
densities set by the Ginzburg length, as it was originally suggested 
\cite{K76} and is still occasionally argued \cite{KR}. 

In this letter we perform the first large-scale numerical studies 
of a temperature quench in 3D $U(1)$ symmetric $\lambda \phi^4$.
This is the familiar Ginzburg-Landau model 
describing the free energy of a neutral system, (eg. $^4He$), 
which as is well known displays a true second order 
phase transition in 3D, with the establishment of long range order
at low temperatures. This implies in particular, contrary to the case of 
lower dimensions, that no defects (vortex strings) can exist in equilibrium  
at sufficiently low temperatures in any causally connected volume \cite{ABH}.

In order to implement the  temperature quench we evolve the fields 
according to:
\begin{eqnarray}
\left( \partial_{t}^2 -\nabla^2 \right) \phi_i - m^2 \phi_i
+ \lambda \phi_i \left( \phi_i^2 + \phi_j^2 \right) 
+  \eta \dot{\phi_i} = \Gamma_i 
\label{e1}
\end{eqnarray}
where $i,j \in \{1,2\}$  and $i \neq j$ in Eq.~(\ref{e1}). 
$\Gamma_i(x)$ is the  Gaussian noise characterized by 
\begin{eqnarray}
\langle \Gamma_i(x) \rangle = 0, 
\qquad  \langle \Gamma_i(x) \Gamma_j(x')\rangle 
= 2 \eta T(t) \delta_{ij} \delta (x-x'),
\label{e2}
\end{eqnarray}
where $x,x'$ denote space-time coordinates. We allow the fields to 
thermalize above the transition and 
proceed to quench the system by changing the noise temperature as
\begin{eqnarray} 
T (t) = T_c - T_0 {t \over t_Q}. 
\label{e3}    
\end{eqnarray}
The time $t_Q$ controls the rate of the quench. 
For $t>{T_c \over T_0} t_Q$, 
$T=0$. 

In the numerical evolution we take the grid spacing  $\Delta x=0.5$
and $\Delta t =0.02$. All results shown are for computational
domains of size $N^3$, with $N=128-160$ and $\eta=1$. 
Strings are detected by integer windings of the field 
phases around lattice plaquettes. 
Vortices hence found are connected by enforcing flux
conservation on each unit volume \cite{ABH}. As a result 
of boundary conditions all strings are closed. 
Above $T_c$, and immediately below strings are
little more than non-perturbative field fluctuations. 
Below $T_c$ they gradually acquire stability and can be 
regarded as either vortex lines or cosmic strings. 

Another useful quantity is the kinetic temperature, $T_K$,
defined as the average kinetic energy per degree of 
freedom. In equilibrium the canonical momentum 
distribution is purely Gaussian 
and $\langle \vert \pi(x) \vert ^2 \rangle  d^{\rm D} x = 2~T$.   
We generalize this for situations away from equilibrium as
\begin{eqnarray}
\langle \vert \pi(x,t) \vert^2 \rangle d^{\rm D} x  \equiv 2~T_K(t).
\end{eqnarray}
All temperature quenches are started at $T_0=1.91 T_c$. 
Even though the choice of initial (high) temperature is somewhat 
arbitrary, it is important that it is 
sufficiently high that the length density in long strings in each 
computational  domain is substantial \cite{ABH}. 

In the immediate vicinity of $T_c$ for a system 
undergoing a second order phase transition, the dynamics of $\phi$ 
are subject to critical slowing down. 
This leads to the estimate of expected density of defects through 
an argument \cite{Z85}, which we briefly reproduce below. 
For the dynamics of  Eqs.~(\ref{e1},\ref{e2})
in the over-damped regime where the 
first time derivative dominates, the characteristic time scale $\tau$
over which the order parameter can react is given by 
\begin{eqnarray}
\tau_{\dot \phi} \simeq  {\eta \over m^2 |\epsilon|^{\nu z}},
\end{eqnarray}
where $\nu$ and $z$ are universal critical exponents, and the 
relative temperature $\epsilon =  {T \over T_c} -1$ 
$= {t \over \tau_Q } \equiv {t \over t_Q} {T_0 \over T_c}$.  
The quench timescale $\tau_Q$ is a 'natural units' rescaling of $t_Q$,
$\tau_Q={T_c \over T_0} t_Q$. 

This over-damped scenario, valid when $\eta^3 \tau_Q > 1$ \cite{LYZ},
is presumably more relevant for condensed matter applications and will 
the the focus of the present Letter. 
Cosmological order parameters may in contrast be under-damped 
-or in reality red-shifted \cite{VS}, 
corresponding to a different dynamics than that 
of Eq.~(\ref{e1},\ref{e2}). 

The characteristic time scale of variation of 
$\epsilon$ is  ${\epsilon \over \dot \epsilon} = t.$
We expect the system to be able to re-adjust to the new equilibrium as
long as the relaxation time is smaller than $t$. Hence, outside the
time interval $\left[- \hat t,\hat t\,\right]$ defined by the equation
$\tau(\epsilon(\hat t)) = \hat t,$
the evolution of $\phi$ is approximately adiabatic, and physical
quantities associated with large length scales will approximately 
follow their (critical) equilibrium values. The time
\begin{eqnarray}
\hat{t}_{\dot \phi} = \pm \left[ 
{\eta \over m^2} \left( 
\tau_Q\right)^{\nu z}  \right]^{1 \over 1+ \nu z}  
; \quad \hat{\epsilon}_{\dot \phi} = \pm \left( { \eta  \over 
m^2} {1 \over  \tau_Q }\right)^{1 \over 1+ \nu z}
\end{eqnarray}
marks the borders between adiabatic and impulse stages of evolution of $\phi$. 
In particular the correlation length $\xi_0$ associated the 
the connected 2-point function above the 
transition will cease to increase as $ \xi = \xi_0 m_0 /|\epsilon|^{\nu}$
once the adiabatic-impulse boundary at $t'=-\hat t$ is reached. 

We expect, then, that the characteristic length scale over which
$\phi$ is ordered already in the course of the transition will be the
correlation length at freeze-out $\hat \xi = 1 / {\hat \epsilon}^\nu$,
\begin{eqnarray}
\hat{\xi}_{\dot \phi} = \left({m^2 \over 
\eta}    \tau_Q \right)^{\nu \over 1+ \nu z}.
\end{eqnarray}
On length scales smaller than $ \hat \xi$ the effect of criticality 
on the field dynamics is negligible.  The initial density of vortex 
lines is then expected to 
scale with $\tau_Q$ as
\begin{eqnarray}
n_{\dot \phi} =  {1 \over (f_{\dot \phi}\hat{\xi}_{\dot \phi})^2} =
{1 \over f_{\dot \phi}^2} \left( { m^2  \over \eta}   \tau_Q  
\right) ^{-\alpha}
\label{e8}
\end{eqnarray}
where $f \sim O(1)$
is a dimensionless factor, parameterizing our ignorance about the exact 
relation between domain size and defect densities, and 
$\alpha={-{2 \nu \over 1 + \nu z}}$. 
Lattice measurements and renormalization group analysis yield $\nu=0.6705$ and 
$z=2.03$ for the very over-damped case when the second time derivative
in Eqs.~(\ref{e1},\ref{e2}) is completely negligible. 
With $\eta=1$ this is not truly the case. We expect $1 \leq z \leq 2.03$,  
implying $ 0.568 \leq \alpha \leq 0.8 $, where the lower and upper 
limit refers to the under-damped case, respectively. 
This is different from mean-field exponents,  $\nu=1/2$ and  
$z=2$, implying $\alpha_{\rm MF}= 0.5$. 
\begin{figure}
\centerline{\psfig{file=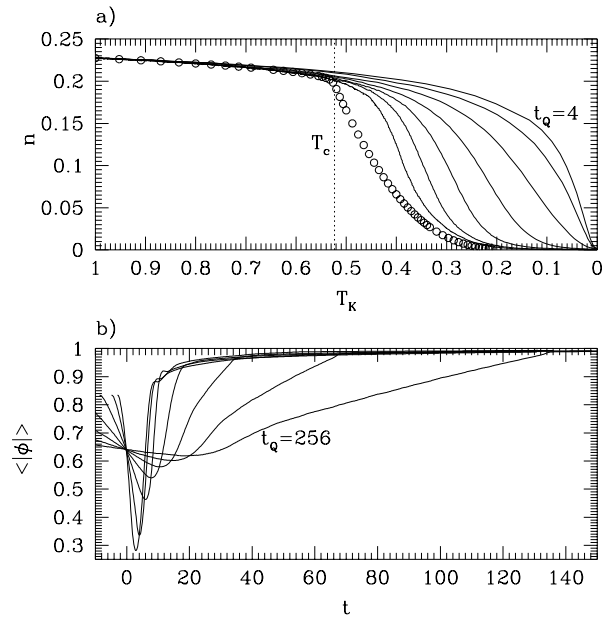,width=3.25in}}
\caption{a) The evolution of the total string density per plaquette $n$ 
with $T_K$. $\circ$ denote thermal equilibrium densities.
b) $\langle \vert \phi \vert \rangle$ vs. t. Solid lines denote 
quench trajectories for $t_Q=4,8,16,32,64,128,256$.}
\label{fig1}
\end{figure}

In a temperature quench the system is evolved 
from an initial state at genuinely high temperature, with a bare 
negative mass squared. As a result the physical mass squared,
$m_{\rm ph}^2$, which takes 
into account the self-energy generated by the high temperature 
field fluctuations is positive and potentially large. 
The temperature quench proceeds by the decrease of  
the external bath temperature at a rate given by $\tau_Q$, 
according to Eq.~(\ref{e3}).
Initially, the system locally
re-thermalizes to the new lower temperature.  
Close to the critical point the physical mass squared  approximately
vanishes leading to the critical slowing down of the field response over 
large spatial scales. This freezes the dynamics of the order parameter:
It can no longer react to the systematic changes of thermodynamic or dynamical
parameters, although slow drift under the combined influence of noise 
and damping continues unabated, even in the large scale 
structure, including long strings. 
Critical slowing down has little effect over the small scale dynamics 
($k^2 >> m_{\rm ph}^2$), which accompanies the externally imposed 
change of bath temperature. 
In this manner critical slowing down sows the seeds for the 
out-of equilibrium dynamics to follow.
Fig.~\ref{fig1}a shows the equilibrium string densities as a function of 
$T_K$, and quench trajectories for several 
values of $\tau_Q$. In 3D $T_K$ is dominated by small scales and   
generally follows faithfully the external bath $T$ variation.

\begin{figure}
\centerline{\psfig{file=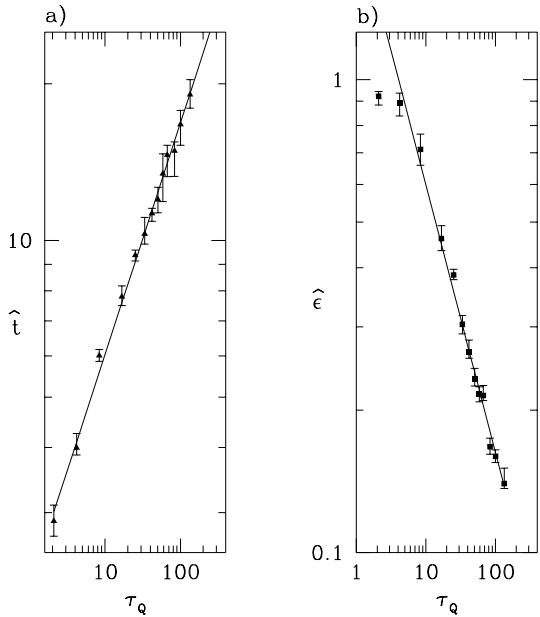,width=3.25in}}
\caption{a) The freeze-out time $\hat t$ as a function of the quench timescale
$\tau_Q$, b) The relative kinetic temperature at $\hat t$, 
$\vert \hat \epsilon \vert = \vert {{T_K(\hat t)} \over {T_c}} - 1 \vert$. 
Slopes of the two lines are $0.445\pm 0.05$  and $0.565\pm 0.016$,
respectively, which compares favorably with theoretically predicted of 
$0.5$. Points corresponding $\tau_Q \leq 8$ exhibit 
saturation, and were ignored in the fitting.}
\label{fig2}
\end{figure}
As predicted by the theory all quenched string densities 
follow the equilibrium trajectory at high temperatures 
but start deviating just above the critical point, 
due to critical slowing down. The magnitude of the effect is
small and difficult to measure. 
After falling out of equilibrium the string densities do not freeze, 
but rather decay slowly while the temperature drops over a period of time. 
This decrease is due mostly to the decay of small scale structure in long 
strings and of small loops.
For faster quenches this regime persists to much lower $T_K$. 
Fig.~\ref{fig1}a clearly shows the hierarchy of string densities 
for different $\tau_Q$'s, at each given $T_K$. 

Fig.~\ref{fig1}b, shows the evolution of the order 
parameter amplitude $\langle \vert \phi \vert \rangle$. 
The persistent cooling of small scale fluctuations  eventually leads  to a 
negative $m_{\rm ph}^2$ which in turn triggers instabilities in the 
long wavelength modes. As these modes  grow (quasi-exponentially)
the $U(1)$ symmetry is spontaneously broken, coarsening the original 
field configuration.
These instabilities are reminiscent of 
those in 'pressure' quenches \cite{LYZ,KR,Dz,Lythe}. 
However here their generation through a negative mass squared 
is explicitly created by the delay in the fields response, 
rather than externally.
The onset of instabilities defines in turn 
$+\hat t$, which we read off from the minima of Fig.~\ref{fig1}b.
Fig.~\ref{fig2}a shows the dependence of ${\hat t}$ on $\tau_Q$.
We can independently confirm the theoretical 
scaling laws, Fig.~\ref{fig2}b by examining the $\tau_Q$ dependence of 
$\hat \epsilon$, computed as the ratio of $T_K$ and $T_c$ at 
the minimum of  $\langle \vert \phi \vert \rangle$.
\begin{figure}
\centerline{\psfig{file=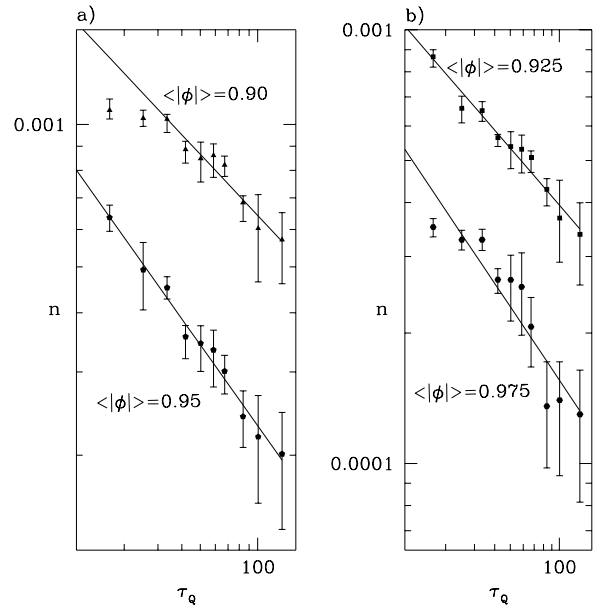,width=3.25in}}
\caption{The string densities measured at 
$\langle \vert \phi \vert\rangle=0.9,0.925,$ $0.95,0.975$. The fits 
are to Eq.~(\ref{e8}), for $\tau_Q \geq 32$,
with $\alpha=0.4296\pm 0.043$, $0.4378 \pm 0.0289$,
$0.5692$ $ \pm 0.0256$, $0.5600$ $ \pm 0.0797$ and
$f_{\dot \phi}=11.11,13.99$, $12.29,15.34$, respectively. The average 
$\bar \alpha =0.4982 \pm 0.079$ and ${\bar f}_{\dot \phi}=13.18 \pm 1.78$.}
\label{fig3}
\end{figure}
Together the results of Fig.~\ref{fig2} confirm the theoretical
scaling  and determine $\nu z=0.82$.
We note however that data corresponding to the fastest quenches appears 
to asymptote to $\vert \hat \epsilon \vert =1$ (and, thus, cease following 
the predicted power law). This is easy to understand, \cite{LYZ}: 
$\hat \epsilon$, 
extracted from the data is the absolute value of the relative temperature 
at the instant when the dynamics is re-started. But, 
$ \epsilon  =  T_K/T_c -1 $ by definition cannot fall to less than $-1$ 
(which happens when $T_K=0$). Hence, when in very fast 
quenches the evolution of $\phi$ re-starts only at $\hat t \geq \tau_Q$, 
$T_c \gg T_K \simeq T(\tau_Q)=0$, and the values of $\vert \hat \epsilon 
\vert$ pile up asymptotically near $1^-$, causing saturation.

At $t=\hat t$ many small scale fluctuations still persist in the
system obscuring the results in terms of string densities. 
It is the subsequent out-of-equilibrium evolution of the fields, 
leading to spontaneous symmetry breaking and ordering, 
that reveals the string densities formed at the quench. 
While the large spatial scales ($k^2 \simeq 0$) 
are unstable and the corresponding modes grow towards 
$\vert \phi \vert=1$, 
the small spatial scales ($k^2 >> m_{ph}^2$) are dissipated away. 
This leads to the emergence of the field configuration created 
by the critical dynamics on large spatial scales, 
and allows the surviving vortex strings to form, i.e. 
to acquire their low energy character of topological defects.
These densities, as a function of $\tau_Q$, are shown 
in Fig.~\ref{fig3}. 
As a criterion to the completion of the transition we measured
$n$ at $t$ such that $\langle \vert \phi \vert \rangle 
= 0.9, 0.925, 0.95, 0.975$. Apart from the cases of saturation, 
the density of strings formed 
follows the theoretical predictions quite satisfactorily, for all 
choices of $\langle \vert \phi \vert \rangle$. The values of 
$f_{\dot \phi} = O(10)$, are similar to 1 and 2D estimates \cite{LYZ}, 
and may be sufficient to explain the non-appearance of vortex lines
in the recent $^4He$ experiment \cite{He4}. 

\begin{figure}
\centerline{\psfig{file=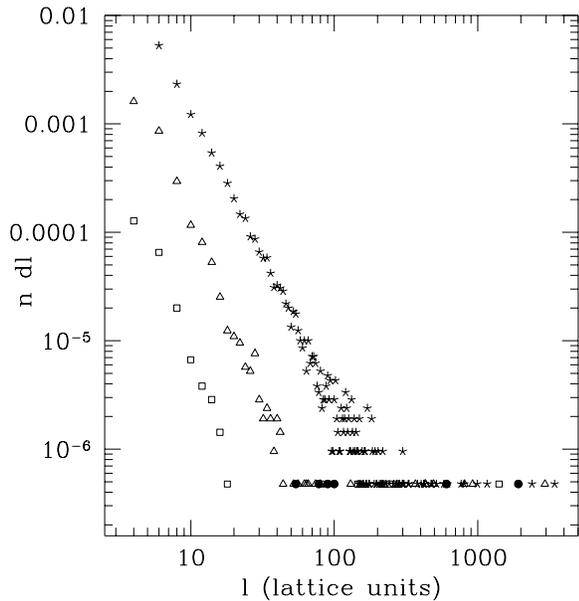,width=3.25in}}
\caption{String length $l$ distributions (n dl vs. l) 
taken between $+ \hat t$ and the 'time of formation' 
($\langle \vert \phi\vert \rangle =0.95$), for $\tau_Q=64$.
Data sets denoted by ($\star$,$\triangle$,$\Box$, $\bullet$) 
correspond to increasingly later times.}
\label{fig4}
\end{figure}
Finally, it is interesting to investigate what kind of strings are formed.
At $T$ above $T_c$ the 'ephemeral string' length distribution 
approaches a Brownian form \cite{ABH}. 
Long random-walk like strings of zeroes of $\phi$ coexist in equilibrium 
with a sea of smaller strongly self-correlated loops. 
As the quench proceeds small scales are dissipated first. 
Nevertheless, the presence of noise and absence of the restoring 
dynamics implies that until $t \simeq + \hat t$ a sizable population 
of small loops can persist. Subsequently the system is dissipated 
further, the symmetry is spontaneously broken and the
fields order starting from the small scales. The result is that by the time 
$\langle \vert \phi \vert \rangle$ has come near its equilibrium 
low-temperature value, 
only long strings, imprinted by the critical dynamics on largest length 
scales, survive, stripped of most of their small scale structure. 
This evolution is shown in Fig.~\ref{fig4}.

Our numerical analysis lends strong support to the general 
picture of dynamical evolution of $\langle \vert \phi \vert \rangle$ 
in a second order phase transition, proposed some time ago 
\cite{Z85}, and partially confirmed 
by experiments \cite{He3} and by simulation of 'pressure' 
quenches in low dimensional systems \cite{LYZ}. The key new aspects of this 
investigation are i) its 3D character, which has allowed us to ii) 
study a to data numerically unexplored temperature quench. In the present 
range of parameters the Ginzburg regime appeared to play no discernible role.
We are currently investigating its effects on the decay of individual strings.

We thank  A.~Gill, T.~Kibble, P.~Laguna, R.~Rivers, 
and A.~Yates for useful discussions. Numerical work 
was done on the T-division/CNLS Avalon Beowulf cluster, 
LANL and  supported by DOE, contract W-7405-ENG-36.

\end{document}